# ERSFQ 8-bit Parallel Binary Shifter for Energy-Efficient Superconducting CPU

A. F. Kirichenko, M. Y. Kamkar, J. Walter, and I. V. Vernik

*Abstract*— We have designed and tested a parallel 8-bit ERSFQ binary shifter that is one of the essential circuits in the design of the energy-efficient superconducting CPU. The binary shifter performs a bi-directional SHIFT instruction of an 8-bit argument. It consists of a bi-direction triple-port shift register controlled by two (left and right) shift pulse generators asynchronously generating a set number of shift pulses. At first clock cycle, an 8-bit word is loaded into the binary shifter and a 3-bit shift argument is loaded into the desired shift-pulse generator. Next, the generator produces the required number of shift SFQ pulses (from 0 to 7) asynchronously, with a repetition rate set by the internal generator delay of ~ 30 ps. These SFQ pulses are applied to the left (positive) or the right (negative) input of the binary shifter. Finally, after the shift operation is completed, the resulting 8-bit word goes to the parallel output. The complete 8-bit ERSFQ binary shifter, consisting of 820 Josephson junctions, was simulated and optimized using PSCAN2. It was fabricated in MIT Lincoln Lab's 10-kA/cm$^2$ SFQ5ee fabrication process with a high-kinetic inductance layer. We have successfully tested the binary shifter at both the LSB-to-MSB and MSB-to-LSB propagation regimes for all eight shift arguments. A single shift operation on a single input word demonstrated operational margins of ±16% of the dc bias current. The correct functionality of the 8-bit ERSFQ binary shifter with the large, exhaustive data pattern was observed within ±10% margins of the dc bias current. In this paper, we describe the design and present the test results for the ERSFQ 8-bit parallel binary shifter.

## I. INTRODUCTION

With energy efficient computing becoming more relevant recently, several projects have been initiated toward building a superconductor-based microprocessor [1].

Despite the requirement for cryogenic cooling and its associated additional energy cost, superconductive electronics has been viewed as an emerging field capable of achieving higher energy efficiency than other technologies [2], [3]. Recently, superconductor-based energy-efficient technologies have been considered as the next generation of digital circuits for computing applications [4]-[9].

We are working on the design and demonstration of an 8-bit ERSFQ [4] CPU with target clock frequency of 10 GHz and energy dissipation below 0.1 nJ per FLOP [10]-[14].

The research is based upon work supported by the Office of the Director of National Intelligence (ODNI), Intelligence Advanced Research Projects Activity (IARPA), via contract W911NF-14-C0090.

All authors are with HYPRES, Elmsford, NY 10523 USA (e-mail: alex@hypres.com).

Fig. 1 shows the block-diagram of the CPU under development. It consists of three major parts: Arithmetic Logic Unit (ALU), Register File, and Instruction memory. In order to fit the CPU on a 5x5-mm$^2$ chip, we have chosen simple, fast, and compact wave-pipeline architecture [11]. The ALU architecture is based on a simple ripple-carry adder with asynchronous carry propagation [12]. This results in a very compact design, but a somewhat limited instruction set. In particular, the ALU lacks a bit shift instruction that is incompatible with the wave-pipeline operation.

A widely implemented device is a barrel shifter [15] using a crossbar approach or a cascade of parallel 2×1 multiplexers. A barrel shifter is a digital circuit that can shift a data word by a specified number of bits with the use of only pure combinational logic, accounting for its wide acceptance by CMOS. A barrel shifter requiring $N \cdot log_2 N$ multiplexers becomes increasingly prohibitive for the hardware of sequential ERSFQ logic. Contrary to CMOS, ERSFQ offers advantages such as inherent high speed and local timing. These can be utilized to implement an energy- and hardware-efficient binary shifter circuit.

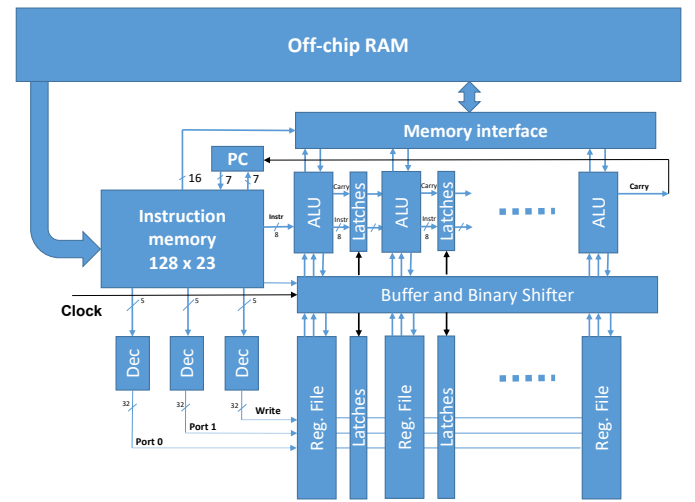

Fig. 1. Block-diagram of the ERSFQ CPU. The binary shifter is between the ALU and the Register File.

In this paper, similar to ARM processor developers [16], we have designed and demonstrated a binary shifter as a separate circuit placed between the ALU and the Register file (Fig. 1). Being limited by on-chip area, and therefore unable to utilize a barrel binary shifter, we have designed and tested an asynchronous 8-bit binary shifter based on a bidirectional shift register (Fig. 2).

This circuit operates asynchronously, producing a shift operation with a latency of 1 to 3 master clock cycles. The maximum latency depends on both the number of bits of the operand and the ratio between the internal (serial) clock of the bit-shifter and the master clock of the CPU.

## II. DESIGN

### A. 8-bit binary shifter architecture

Fig. 2a shows a block-diagram of the 8-bit binary shifter. The binary shifter comprises a bidirectional shift register (Fig. 2b) integrated with two fast clock generators (for left and right shift operation).

The shift register has a parallel 8-bit operand input and a parallel 8-bit result output. The operation starts with loading the first operand (an 8-bit number to shift) into the input of the shift register, while simultaneously loading a 3-bit shift operand to one of the two clock generators (to instruct the number of bits to shift by). A *"launch"* signal is then triggered to start the clock generator. Every SFQ pulse produced by the left clock generator makes a 1-bit shift of the input operand to the left (and the same goes for the right-side clock generator shifting bits to the right). The clock generator is designed to asynchronously generate $N$ SFQ pulses, where $N$ is the shift operand loaded into the clock generator, at the rate of ~30 GHz for the 10-kA/cm$^2$ fabrication process [17]. After generating clock pulses, the generator issues a parallel readout signal to the shift register. Thus, the whole shift operation is accomplished in a single clock cycle, similar to a barrel binary shifter [15].

The whole 8-bit binary shifter was simulated and optimized for margins in excess of +/-25% in PSCAN2 physical-level simulator [18].

### B. Bidirectional shift register

The bidirectional 8-bit binary shift register is constructed of triple-port destructive readout flip-flops (D3) [19]. This cell is just a simple modification of the dual-port (D2) flip-flop [20]. Fig. 3b shows the schematics of this cell. The cell has an input *Set*, as with any D flip-flop, and three destructive readout ports (*In1-In3*) with corresponding outputs (*O1-O3*).

As shown in Fig. 2b, D3 cells are connected in shift register via port 1 for the right shift (*SR*) and via port 2 for the left shift (*SL*). The third port is used for the result readout (*Read*).

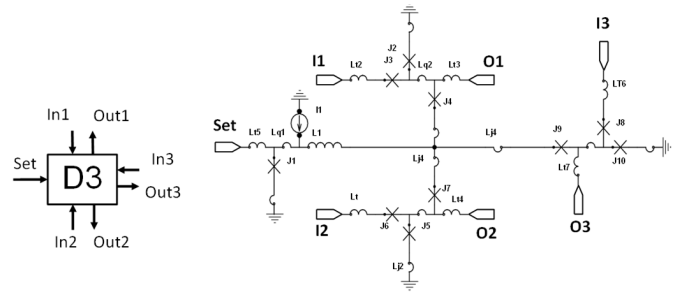

Fig. 3. Notation (a) and schematics (b) of D3 flip-flop

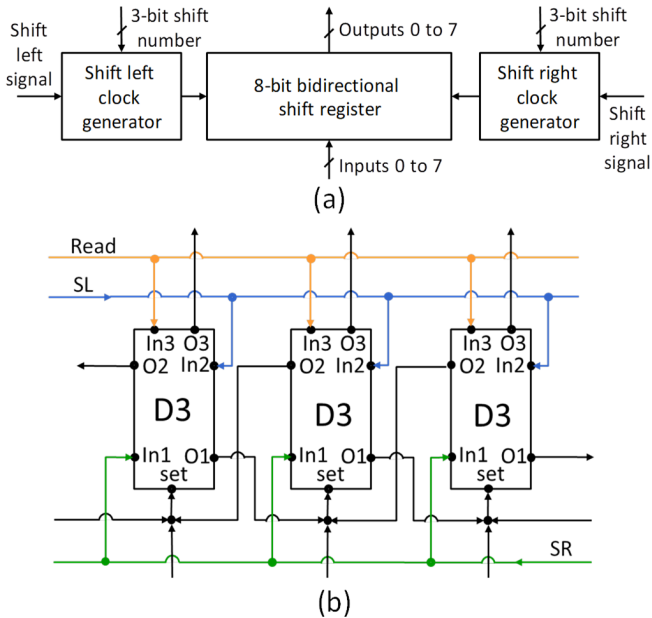

Fig. 2. Block diagrams of the 8-bit binary shifter (a) based on a bidirectional shift register (b).

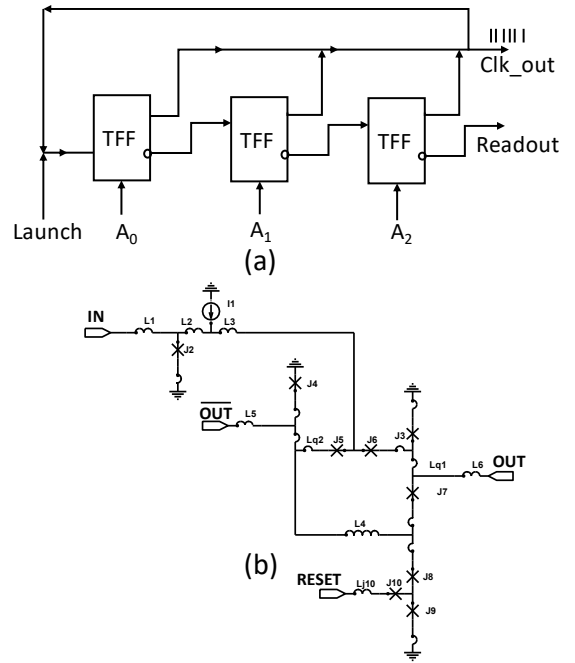

Fig. 4. Block diagrams of the clock generator (a) and the TFF as a resettable T flip-flop (b).

*C. 3-bit clock generator*

The clock generator (Fig. 4a) comprises a chain of three resettable T flip-flops (Fig. 4b) connected in a loop. The direct output of each TFF goes to the feedback loop connecting the input of the first TFF with all TFF outputs. The inverted output of each TFF is connected to the input of the following TFF. The inverted output of the final TFF produces the *Readout* signal. Thus, as it can be seen in Fig. 4a, every *Launch* signal produces a number of SFQ pulses that corresponds to the initial state of TFFs. The initial state of the counter is set by a 3-bit input (*A0-A2*). The actual number of generated pulses is the inverted 3-bit operand A.

*D. Bias line and Feeding JTL*

Maintaining ERSFQ biasing scheme is a bit challenging in this circuit because it operates at a different speed than the CPU clock frequency. The asynchronous clock generators are generating serial clock at 8-times higher rate than the master clock of the CPU. Consequently, the bit-shifter must be connected to a separate bias line.

The clock generator itself is made of T flip-flops operating at different frequencies. To mitigate possible bias current redistribution, this circuit requires a feeding JTL with bias current no less than 1/4 of the total dc bias current of the bit-shifter. In the design, we have placed such a JTL under the bidirectional shift register (Fig. 5).

### III. Layout

The 8-bit ERSFQ binary shifter was fabricated with the MIT-LL SFQ5ee 10 kA/cm$^2$ fabrication process [17] featuring eight Nb wiring layers and a high kinetic inductance layer (HKIL). The HKIL allows placement of large ERSFQ bias inductors under the cells reducing the size of an ERSFQ circuit.

Fig. 5 shows the layout of the 8-bit ERSFQ binary shifter with all components (such as bidirectional shift register cell and clock generators), with all inputs and outputs marked. The inset shows the entire 5 x 5 mm$^2$ test chip. The whole circuit utilizes a total of 820 Josephson junctions (including ERSFQ bias JJs) and occupies an area of 1.1 x 0.13 mm$^2$. The binary shifter has a 90-μm bit slice pitch.

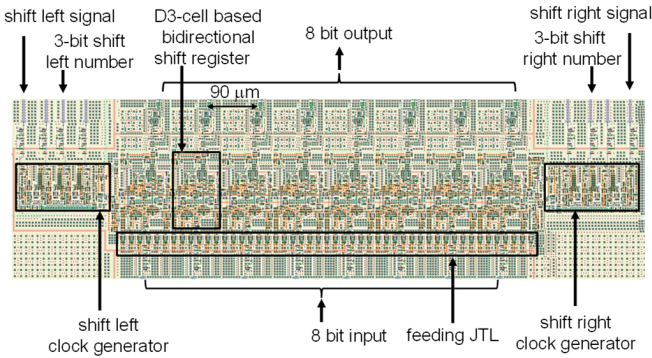

Fig. 5. Layout of 8-bit ERSFQ bidirectional bit shifter with all components marked. It is designed for MIT LL SFQ5ee 10 kA/cm$^2$ process with a high-kinetic inductance layer.

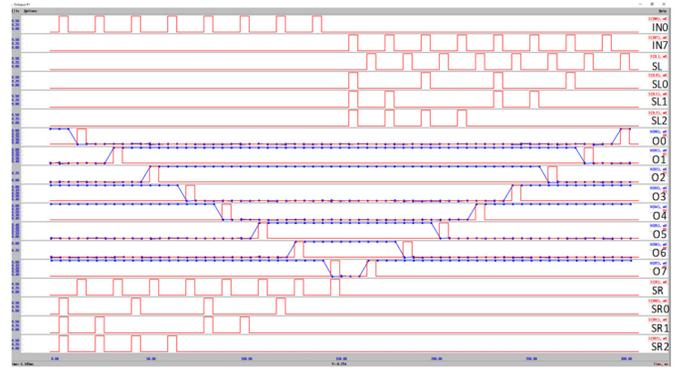

Fig. 6. Correct operation of 8-binary shifter with bias margins of ±10% (bias from 106 mA to 129 mA with ERSFQ $I_C$ = 121 mA). The red traces indicate input and digitized output signals. The blue traces are the measured standard toggling SFQ/dc converters outputs (*O0-O7*).

### IV. Experimental Results

Our tests have successfully demonstrated the operation of the 8-bit ERSFQ binary shifter. The exhaustive evaluation of functionality and the bias margin measurements were performed with the aid of the OCTOPUX automated test system [21].

Testing of the binary shifter was performed as follows: First, two arguments were loaded—an 8-bit number into the parallel input and an inverted 3-bit shift argument into either the shift-left or shift-right clock generator. Next, a launch signal was applied to the selected generator, triggering the production of the prescribed number of SFQ shift pulses (from 0 to 7) that were then transmitted to the corresponding left or right input of the binary shifter. These pulses are asynchronously produced on-chip, with a repetition rate set by the internal generator delay (see Fig. 4a) of ~ 30 ps. Finally, after the shift operation was completed, the resulting 8-bit word was read out at the parallel output with standard toggling SFQ/dc converters.

We have successfully tested the binary shifter for each of the possible eight shift arguments, in both directions, with various 8-bit inputs. A single shift operation on a single operand functioned within margins of ±16% of the dc bias current.

Fig. 6 shows the correct functionality of the 8-bit ERSFQ binary shifter with the most exhaustive test pattern. It consists of two halves: first a right shift of the LSB by 0 to 7 bits, and then a left shift of the MSB by 0 to 7 bits.

The left half in Fig. 6 comprises eight right shift operations of the LSB (*IN0*) by eight different shift values defined by the bit-wise inverted signals on *SR0, SR1, and SR2*. Subsequently, "shift right" launch signal (*SR*) produces the resulting output *(O0-O7)*. Therefore, the first operation on the pattern shows the LSB right-shifting by 0 (inverted 7) and producing an output on *O0*. The next operation shows the LSB right-shifting by 1 position (inverted 6) and producing an output on *O1*. This continues until the final right-shift of the LSB by 7 (inverted 0) produces an output on *O7*.

Similarly, the right half of the pattern demonstrates the left-shift operation on the MSB (*IN7*). In the same manner, the bit on *IN7* is left-shifted sequentially by 0, 1, 2, etc., producing outputs on *O7, O6, O5,* etc.

The measured "staircase" output pattern confirms the correct operation for all possible 16 shifts observed on all outputs from *O0* to *O7*. The correct operation has been confirmed for this exhaustive pattern within ±10% dc bias current margins (from 106 mA to 129 mA). The zero-static power dissipation operation [22] was observed within ±3% bias margins (bias current held below the critical current of the ERSFQ bias network $I_C$ = 121 mA). The reduced bias margins as compared to the simulation are caused by fabrication issues affecting bias current delivery via bias inductors formed by the high-kinetic-inductance layer (HKIL). If a particular HKIL bias inductor was defective, causing bias delivery to the ERSFQ cell to fail, the circuit would still work under overbiased conditions with the required bias current redistributed from the adjacent cells.

## V. Conclusion

We have designed and tested a parallel 8-bit ERSFQ binary shifter. The compact design based on bi-direction triple-port shift register enables us to fit the binary shifter into a relatively small 1.1 x 0.13 mm² area with a 90 μm slice pitch to allow integration into an ERSFQ CPU on a 5 x 5 mm² chip.

The ERSFQ bit-shifter circuit comprises 820 Josephson junctions. It was simulated and optimized in PSCAN2 physical-level simulator and fabricated using the 10-kA/cm² MIT-LL SFQ5ee fabrication process with eight Nb wiring layers and HKIL.

The 8-bit bit shifter has been demonstrated at a low master clock frequency while the internal serial shift signal rate was about 30 GHz. This produces a latency of ~300 ps (three 10-GHz master clock cycles).

The correct functionality of the binary shifter with the exhaustive data pattern was observed within ±10% dc bias margins.

## Acknowledgment

The authors would like to thank the MIT Lincoln Lab foundry team for circuit fabrication and S. Holmes and M. Manheimer for support and encouragement.

The views and conclusions contained herein are those of the authors and should not be interpreted as necessarily representing the official policies or endorsements, either expressed or implied, of the ODNI, IARPA, or the U.S. Government.